\documentstyle[aps,prl,epsfig,amsfonts,amsmath,array,cite,citesort,float,preprint]{revtex}
\restylefloat{figure}

\def\sr14{Sr$_{14}$Cu$_{24}$O$_{41}$\ }

\begin{document}

\title{Exact bounds on the critical frustration in the Shastry--Sutherland Model }
\author{Ute L\"ow and Erwin M\"uller-Hartmann}
\address{Institut f\"ur Theoretische Physik, Universit\"at zu K\"oln}
\date{\today}
\maketitle
\begin{abstract}
We discuss the phase diagram of the Shastry--Sutherland model
for arbitrary spin $S$ and derive rigorous lower and upper bounds on
the phase boundaries of the dimer phase by using various versions 
of a variational ansatz in combination with the exact diagonalisation
method.

\end{abstract}
\hspace{1.9 cm}

\section{Introduction}
\label{sec:Intro}

\begin{figure}[ht]
\begin{center}
\epsfig{file=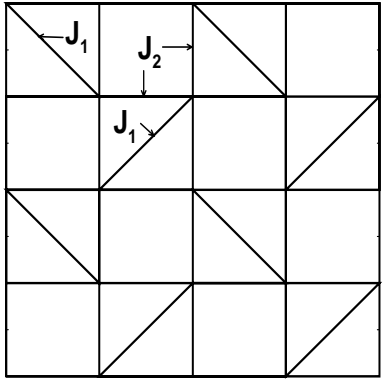,width=8cm}
\end{center}
\caption{Shastry--Sutherland lattice with 
coupling $J_1$
along the diagonal and $J_2$ along the vertical and horizontal bonds.}
\label{fig0}
\end{figure}

We study  a two-dimensional Heisenberg model with 
additional frustrating interactions of strength $J_1$ on 
every second diagonal
bond (see Fig.~\ref{fig0}) given by the Hamiltonian

\begin{equation}
  \label{eq:hamilton}
  H= J_2 \sum_{<i,j>} {\bf S_i S_j}
  +J_1  \sum_{<i,j>Dimer}{\bf S_iS_j},
\end{equation}
where $\bf S_i$ denotes a spin operator for spin $S$ at site i.
The model, commonly referred to as the Shastry--Sutherland 
model \cite{SS}, displays 
a rich zero-temperature phase diagram as a function of $J_1$ and $J_2$,
showing long range ferromagnetic, antiferromagnetic and helical order,
as well as short range spin liquid behaviour.
Also for large frustrating coupling  $J_1$ 
the model has an exactly known ground state,
built up of uncoupled dimers on the diagonal bonds.
It was the existence of this exact 
ground state, which has first attracted 
attention to the model.

Recently the Hamiltonian~(\ref{eq:hamilton}) for  $S=\frac{1}{2}$
has been widely discussed, because it 
represents essential features of SrCu$_2$(BO$_3$)$_2$, 
a newly discovered spin gap substance 
\cite{Kageyama}.
SrCu$_2$(BO$_3$)$_2$ appears to have a ground state 
of dimer singlets and
some of its unusual magnetic properties are well explained by 
the Shastry--Sutherland Hamiltonian.

\begin{figure}[!htb]
\begin{center}
\epsfig{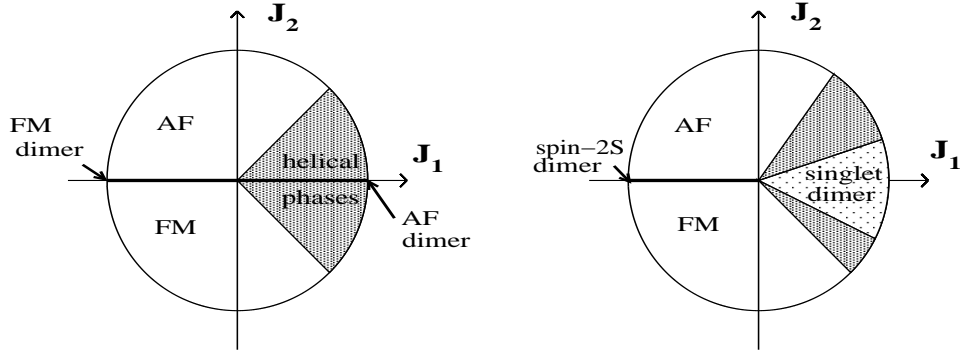}
\end{center}
\vskip 2.0cm
\caption{Exact phase diagram of the classical Heisenberg model (left) and schematic phase diagram of the Heisenberg model for $S>\frac{1}{2}$ (right). 
For $S>\frac{1}{2}$ adjacent to the dimer phase two phases 
occur, whose nature is still a point of controversy.} 
\label{fig1}
\end{figure}

In the classical limit ($S=\infty$) of the Hamiltonian~(\ref{eq:hamilton}) 
the nature of
the ground state phases and their boundaries are easily analyzed 
(see Fig.~\ref{fig1}). 
The model shows two long range helical phases
for $0<|J_2|<J_1$, which are separated by an antiferromagnetic
dimer phase along the line $0=J_2<J_1$.
In the regime $|J_2|>J_1$ the ground state is ordered  
antiferromagnetically for $J_2>0$ and ferromagnetically 
for $J_2<0$. For $J_2=0>J_1$ a phase of independent
ferromagnetic dimers occurs.

The ground state energy of the helical phase can be found
by minimizing the energy
of a cluster of only three spins. From this basic 
three spin entity  the ground state of
the whole system is constructed. Depending on the initial 
choice of spin directions
one finds a spin helix with a four-fold degenerate direction
and with a twist angle of $\theta=\pi\pm\arccos(J_2/J_1)$ 
between neighbouring spins\cite{AM}.

Away from the classical limit quantum effects 
transmute the classical antiferromagnetic dimers for $J_1> 0$ into
unique singlet dimers and
stabilize the resulting
dimer phase.
In what follows we will often use  the variable
$x:=J_2/J_1$, which is the inverse  
of the frustration. For $S<\infty$ and $J_2>0$ the dimer phase occupies a 
finite region of the phase diagram and 
at a certain critical value $x_c^a(S)> 0$ 
there is a phase transition 
of first order \cite{AM} to 
a  new phase, whose nature is controversially discussed at the moment.
For even larger $x$ the system
changes to the antiferromagnetic regime.

In the ferromagnetic region ($J_2<0$)
the situation is slightly simpler, since 
for $S=\frac{1}{2}$ there is a first order phase transition 
directly from the dimer to the ferromagnetic phase 
at $x=-1$. This transition marks one of the 
exactly known points in the phase diagram.
For~$S>\frac{1}{2}$,
like in the antiferromagnetic regime,
the system crosses at a negative $x_c^f(S)$ from the dimer 
to an intervening phase.
Finally at $x=-1$ the ferromagnetic region is reached for all $S$.

For $J_1<0$ the classical phase diagram is not changed by quantum
effects. The ferromagnetic dimer phase for $J_2=0$
becomes a phase of independent spin-$2S$ dimers with 
macroscopic degeneracy. It may be interesting to note, that 
the spin-$S$ square lattice antiferromagnet 
for $0=J_1<J_2$ is
continuously connected to the spin-$2S$ 
square lattice antiferromagnet as $J_2> 0 >J_1\to -\infty$.

There are two major problems connected with the
phase transition between the short range spin liquid phase
and the adjacent phases.
Firstly the nature of the phases adjacent to the singlet dimer phase 
- both on the ferromagnetic and on the antiferromagnetic side -
is not clear and a point of intense investigation at the moment.
Secondly for arbitrary $S$ the boundary 
of the dimer phase cannot
be determined by a simple argument like in the case $S=\infty$.

In this paper we concentrate on the second problem
and give rigorous lower (Sec.~\ref{low})  and upper (Sec.~\ref{up})
bounds on the phase boundaries of the dimer phase by 
using various versions of a variational ansatz for finite clusters.

Various approximative methods have been used recently 
to determine this phase boundary.
In Ref.\cite{AM} the phase boundary was calculated by means of
Schwinger boson mean-field theory.
By the flow equation method
in Ref.\cite{KBMU} a value of $x_c^a(S=\frac{1}{2})=0.63$ was obtained
and in Ref.\cite{KK}
an intervening plaquette phase was suggested for $S=\frac{1}{2}$
occupying the regime $0.861> x >0.677$.
In Ref.\cite{CMS} by an extension of $SU(2)$ to the symplectic groups
$Sp(2N)$ an intervening phase with helical and incommensurate order
was found between a dimer phase and a region with collinear commensurate
order. Also in Ref.\cite{CB} 
it was found 
for a two-dimensional model with frustration,
that the dimer and the antiferromagnetic phase are separated 
by an intervening regime, which is characterized as a weakly incommensurate 
spin density wave.

\section{Lower bounds on ${\lowercase{x}}_{\lowercase{c}}^{\lowercase{a}}$
and on ${-\lowercase{x}}_{\lowercase{c}}^{\lowercase{f}}$ }
\label{low}

\begin{figure}[!htb]
\begin{center}
\epsfig{file=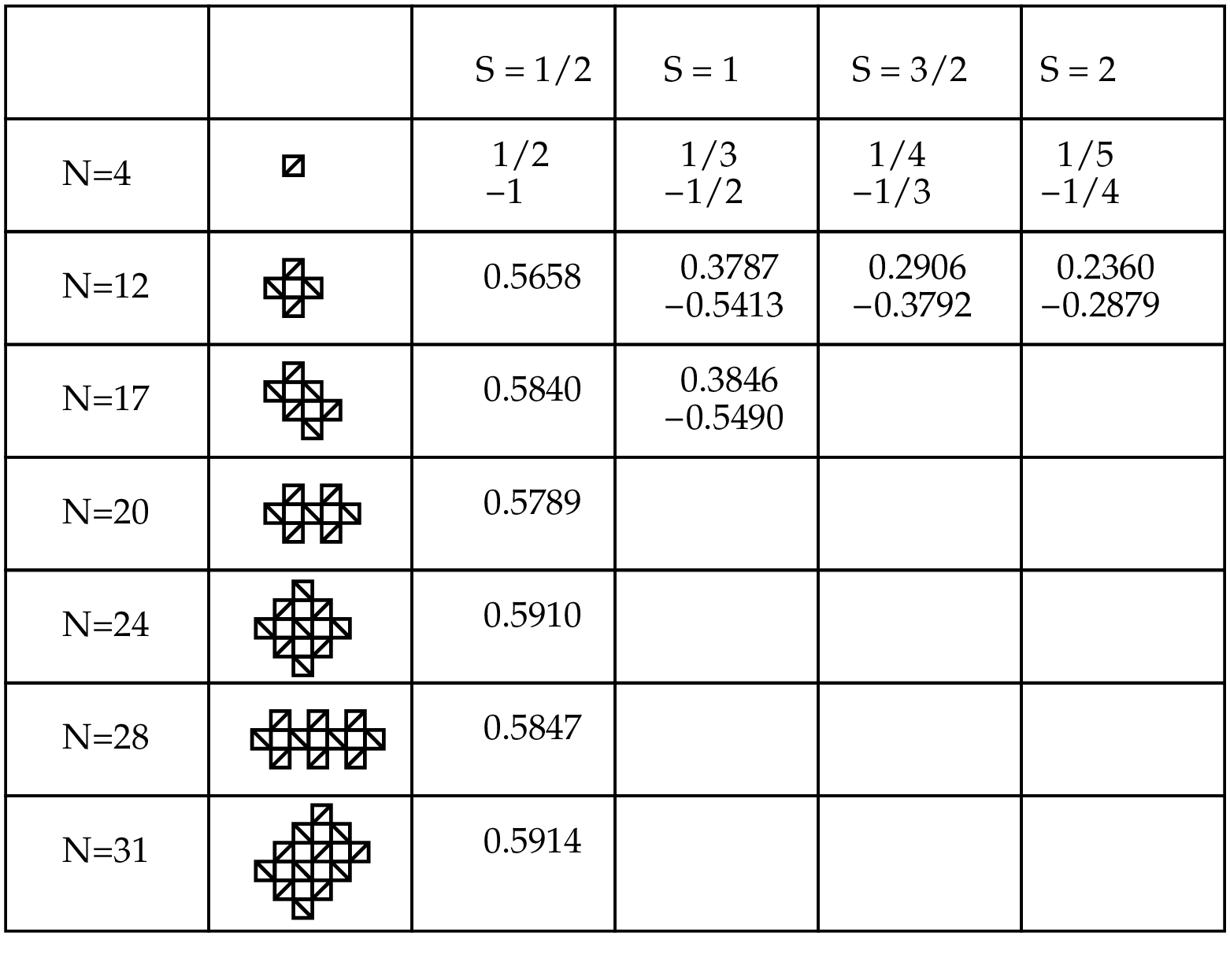,width=8cm}
\end{center}
\caption{$x_c^a$ for $S=\frac{1}{2},1,\frac{3}{2},2$ 
calculated on finite clusters.
The negative numbers denote $x_c^f$ for $S=1,\frac{3}{2},2$.}
\label{fig2}
\end{figure}

To obtain lower bounds on $x_c^a$ we decompose the Hamiltonian $H$ 
of Eq.~(\ref{eq:hamilton}) into cluster terms $H_i^N$ with
\begin{equation}
  \label{eq:sum}
  H=  \sum_i H_i^N
\end{equation}
and calculate the lowest eigenvalues $E_N^0$ 
of the Hamiltonian $H_i^N$ of the clusters 
(N is the number of spins in the cluster).
The decomposition of $H$ is chosen in such a way, that 
the clusters cover the whole lattice without overlapping bonds.
Taking the ground state of H as variational state for
the finite clusters \cite{Anderson} it follows 
that $E_N^0$ is always smaller than or equal to the ground state energy
$E^0_\infty$ of $H$.  To obtain a bound on $x_c^a$
we choose finite clusters (see Fig.~\ref{fig2})
which have the dimer state as an eigenstate 
and  calculate their critical value
$x_{c,N}^{a}$. It is obvious, that
$x_{c,N}^{a}$ is
always smaller than the $x_c^a$ of the infinite system.

As the simplest possible system 
we consider a plaquette with four spins
(first entry in Fig.~\ref{fig2}).
In this case the energy of the second lowest 
state, which for growing x is competing with the dimer state, crosses the
energy of the dimer ground state 
at

\begin{equation}
  \label{eq:low_4}
 x_{c,4}^{a}(S)=\frac{1}{2S+1}.
\end{equation}
From what was said above  $x_{c,4}^{a}(S)$
is a strict lower bound for $x_c^a$.
(Note that for $S>1/2$ the first estimate Eq.~(\ref{eq:low_4}) is better
than the bound $J_1/J_2>2(1+S)$ given in Ref.\cite{AM}.)
Better limits on $x_c^a$  were obtained by calculating 
the ground state energies of the systems
shown in Fig.~\ref{fig2} using the Lanczos method.

For a system with 31 sites we thus obtain  a best lower bound
of 0.5914 for $x_c(S=\frac{1}{2})$.
We want to point out, that even for $S=\frac{1}{2}$, where systems
of up to 31 spins were calculated, it is difficult to make a good finite
size analysis with the results of the finite clusters,
because not only the system size, but also the shape of the cluster 
influences the ground state energy. 
On these grounds 
the significance of the recent estimate
of $x_c(S=\frac{1}{2})=0.7\pm 0.01$ in Ref.\cite{MU},
which was obtained by an extrapolation of three systems 
with {\sl different shapes}, appears questionable.

\begin{figure}[!htb]
\begin{center}
\epsfig{file=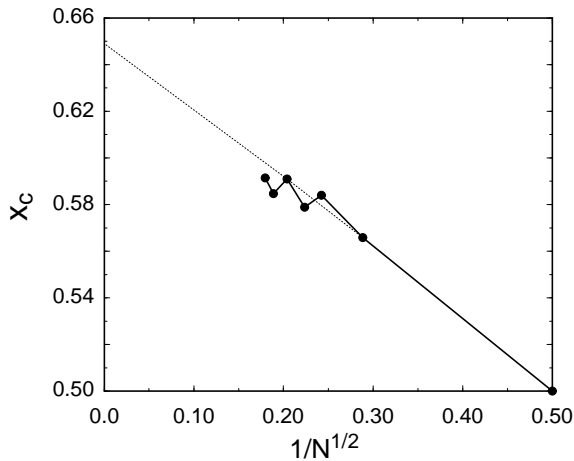,width=8cm}
\end{center}
\caption{$x_c^a(S=\frac{1}{2})$ for the systems shown in 
Fig.~\ref{fig2} plotted versus $1/\sqrt{N}$. The dotted line connects the
results for $N=12,24$ sites, but should not 
be taken as an exact lower bound.}
\label{fig3}
\end{figure}

To show the tendency of a possible 
extrapolation we  plot in Fig.~\ref{fig3} 
the results for $x_c$ for  $S=\frac{1}{2}$ as a function 
of $1/\sqrt{N}$  and connect by a straight line the results 
of  systems with $N=12,24$ spins which have corresponding shapes.
 The point $x_c=0.65$, where this line meets the 
ordinate shows a reasonable value for $x_c$, but obviously cannot be taken as 
a serious extrapolation.
The  best bounds we obtained for $S=1,\frac{3}{2},2$ 
are given in Fig.~\ref{fig2}.

Let us now consider the ferromagnetic regime. 
We find, that the dimer state of a  four spin plaquette 
is the ground state for $x \geq -\frac{1}{2S}$ which implies

\begin{equation}
  \label{eq:low_ferr}
 x_c^f(S) \leq -\frac{1}{2S}.
\end{equation}
This bound is exact for $S=\frac{1}{2}$, since $x_c^f=-1$
coincides with the boundary of the ferromagnetic phase.
For $S> \frac{1}{2}$ we again find an improvement on the result
of the four spin system 
by considering larger clusters. Some results for $S=1,\frac{3}{2},2$ 
are shown in Fig.~\ref{fig2}.

\section{Upper bounds on ${\lowercase{x}}_{\lowercase{c}}^{\lowercase{a}}$
and on ${-\lowercase{x}}_{\lowercase{c}}^{\lowercase{f}}$ }
\label{up}

Upper bounds for the stability of the dimer phase 
can be obtained by considering a variational ansatz 
for finite clusters. We only briefly sketch the idea here,
since it is widely used in the literature
(see e.g. Ref.\cite{WS} and references therein).

\begin{figure}[!htb]
\begin{center}
\epsfig{file=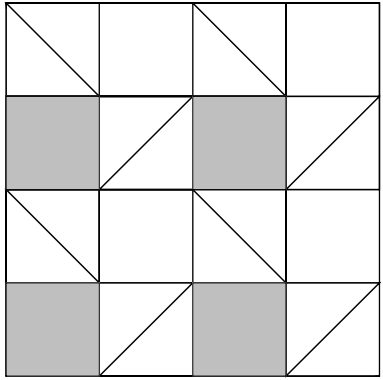,width=8cm}
\end{center}
\caption{Four site clusters (shaded grey) surrounded by 
external bonds. The product state built up of the 
eigenstates of the cluster Hamiltonian 
is used as variational ansatz to obtain an upper bound on $x_c^a$.}
\label{fig4}
\end{figure}

The Hamiltonian is split into clusters without 
common spins and external bonds connecting the clusters,
\begin{equation}
  \label{eq:upper}
 H=\sum_{cluster} H_{cluster}^N +H_{bond},
\end{equation}
as indicated for four site clusters in
Fig.~\ref{fig4}.  We use as variational state 
$\prod_{cluster} |\Psi_{cl}\rangle$, 
where $|\Psi_{cl}\rangle$  is the ground state of $H_{cluster}^N$. 
If the total spin of the clusters is zero in the ground state
the expectation value of the external bonds vanishes

\begin{equation}
\langle \prod_{cluster} \Psi_{cl}|H_{bond}| \prod_{cluster} \Psi_{cl}\rangle=0.
\end{equation}
We have calculated the ground state energies $E^N$ of $H_{cluster}^N$
for clusters with up to N=32 spins 
and obtained upper bounds on $x_c^a$
by comparing them with the energy of the dimer state.

\begin{figure}[!htb]
\begin{center}
\epsfig{file=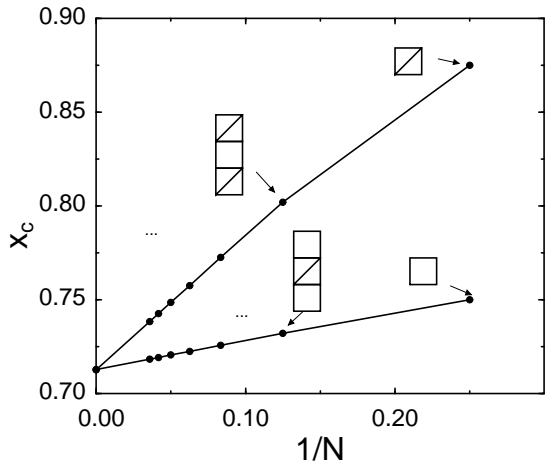,width=8cm}
\end{center}
\caption{ Extrapolation of the upper bounds obtained from 
stripe like configurations. Linear extrapolation of the last two points yields
0.7126 and 0.7127.}
\label{fig5}
\end{figure}

\begin{figure}[!htb]
\begin{center}
\epsfig{file=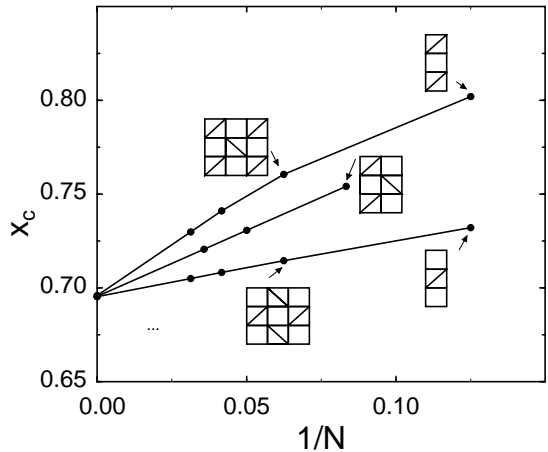,width=8cm}
\end{center}
\caption{Same extrapolation as in Fig.~\ref{fig5} for stripes of width four. Linear extrapolation of the last two points yields 0.6959, 0.6955 and 0.6953 for the three different shapes.}
\label{fig6}
\end{figure}

For a four spin plaquette and arbitrary spin $S$ we thus find 
\begin{equation}
  \label{eq:upper1}
x_c^a \leq \frac{S+1}{2S+1}.
\end{equation}
For stripe configurations of the type  $2\times \frac{N}{2}$ 
the resulting $x_c^a$ for $S=\frac{1}{2}$ 
are shown  as a function of one over 
system size in Fig.~\ref{fig5}. 
Note that results for two different types of stripes 
(as indicated in the figure) are given and that both types 
extrapolate to almost the same point 0.7126(1) for $N \rightarrow \infty$.
A similar extrapolation for stripes of width four 
(see. Fig.~\ref{fig6}) gives an upper bound of
$x_c^a=0.6955(4)$ for $S=\frac{1}{2}$, which is better than
the bound obtained for stripes of width two.
Again three different shapes converge to about the same result as N approaches
infinity.
The lowest value in Fig.~\ref{fig6}
was obtained for a $N=32$ system
yielding the exact bound $x_c^a \leq 0.7050$.
An extrapolation of stripes of width two for S=1 
and systems of up to 16 spins gives 
a best upper bound of $x_c^a(S=1)\leq 0.618$.

In the limit $S \rightarrow \infty$ 
the upper bound (\ref{eq:upper1}) derived from the four spin system 
does not provide
the correct behaviour expected for the classical Heisenberg model.
We can however find another bound, which yields a 
better upper limit 
for $S>1$ by using a helical
product state as a variational state.

The ground state energy for $S=\infty$ in the helical phase 
\cite{AM} is given by

\begin{equation}
  \label{eq:hel1}
E_{hel}=-\frac{|J_1|}{2}-\frac{J_2^2}{J_1}
\end{equation}
where the angle between two neighbouring spins 
is   $\theta=\pi\pm\arccos(J_2/J_1)$ 
and the length of the spins is normalized to 1.

We consider as variational ansatz a system of (quantum) 
spins polarized 
along the directions of the classical spins.
Comparing the expectation  value of this state
with the energy of the dimer state 
we find

\begin{equation}
  \label{eq:hel}
-\left(\frac{|J_1|}{2}+\frac{J_2^2}{J_1}  \right) S^2 < -J_1 S(S+1)
\end{equation}
from which 
\begin{equation}
  \label{eq:hel_bound}
x_c^a < \frac{1}{\sqrt{2S}}
\end{equation}
results as a criterion for the instability of the
dimer state.
Thus for $x> 1/\sqrt{2S}$ 
the dimer state is no longer the lowest state, because
the helical product state has  a lower energy.
Since the ground state energy of the helical state does not 
depend on the sign of $J_2$ 
Eq.~(\ref{eq:hel_bound})  also gives an upper bound on
$-x_c^f$ in the ferromagnetic regime.

\section{Summary}

In this paper we have discussed the phase diagram of the Shastry--Sutherland 
model for arbitrary spin $S$. In the regime $J_1<|J_2|$ the phase diagram for 
the quantum models ($S<\infty$) fully agrees with that of the classical model
($S=\infty$). Quantum effects do, however, modify the classical phase diagram
in the regime $0<|J_2|<J_1$. As emphasized already by Shastry and Sutherland 
\cite{SS} the dimer phase is strongly stabilized by quantum effects and
exists in a regime $x^f_c(S)<J_2/J_1<x^a_c(S)$ which grows with decreasing spin
S. The nature of the intervening phases originating from the classical helical
phases is not totally clear, but their regime shrinks as the dimer phase 
expands. For $J_2<0$ the intervening phase between the dimer and the 
ferromagnetic phase is certain to exist for $S>1/2$, but completely disappears
for $S=1/2$. For $J_2>0$ exact statements on the existence of an intervening 
phase between the dimer and the N\'eel phase are not available, 
but in Ref.\cite{KK} 
rather strong evidence in favor of its existence for $S=1/2$ is provided. 

\vskip 2.2cm
\begin{figure}[!htb]
\begin{center}
\epsfig{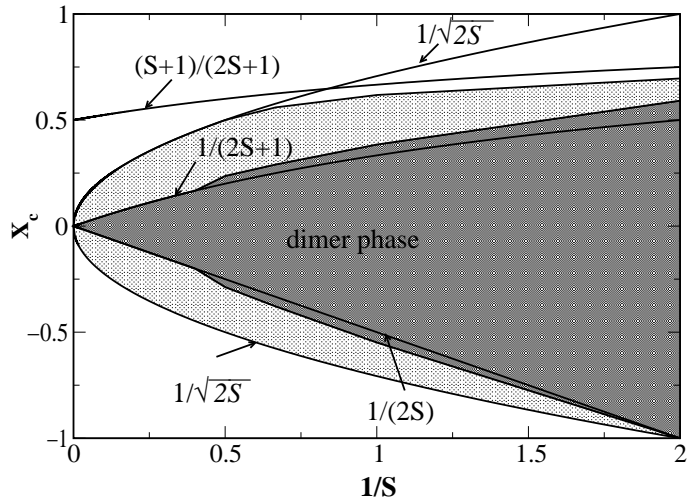}
\end{center}
\vskip 1.5cm
\caption{Summary of bounds for the dimer phase in
the Shastry-Sutherland model.}
\label{fig7}
\end{figure}

The emphasis of the present work was put on the derivation of exact upper and 
lower bounds on the boundaries $x^f_c(S)$ and $x^a_c(S)$ of the dimer phase. 
Our results on this topic are summarized 
in  Fig.~\ref{fig7}. The dark grey area shows 
the region where the dimer state is certain to be the ground state and the 
white area is the region where the dimer state is certainly 
not the ground state. From the results presented in this paper 
it follows, that the exact boundary must be located in the light 
grey area.
The black lines in Fig.~\ref{fig7} represent curves 
discussed in the text.
The upper bounds for $x_c^a(S)$ were obtained by 
a helical variational ansatz ($\frac{1}{\sqrt{2S}}$), by a finite 
cluster ansatz ($\frac{S+1}{2S+1}$) and for $S\leq 1 $ 
by extrapolating results for series of finite clusters
(see Figs.~\ref{fig5} and ~\ref{fig6}).
The lower bounds on $x_c^a(S)$ were derived from four spin plaquettes 
($\frac{1}{2S+1}$) and from the finite systems shown in Fig.~\ref{fig2}
for $S\leq 2$.
In the ferromagnetic regime we have a lower bound 
($\frac{-1}{\sqrt{2S}}$) 
from the helical ansatz, 
which is 
shown together with the upper bounds obtained from 
the largest tractable finite clusters (Fig.~\ref{fig2}).
The intervening phase on the ferromagnetic 
side extends down to $x=-1$ (bottom of Fig.~\ref{fig7})
where the ferromagnetic regime begins for all $S$.

\section{Acknowledgments}
The authors gratefully appreciate useful
discussions with K.~Fabricius, G.~Uhrig, and J.~Zittartz.

\end{document}